\documentclass[12pt,preprint]{aastex}
\usepackage{natbib,epsfig,wrapfig,floatflt,ulem}
\def\h2o{H$_2$O}
\def\ch4{CH$_4$}
\def\arcs{\ifmmode {''}\else $''$\fi}

\shorttitle{The Lowest-Mass Member of $\beta$ Pictoris Moving Group}
\shortauthors{Rice et al.}

\begin{document}

\title{The Lowest Mass Member of the $\beta$ Pictoris Moving Group}

\author{{Emily L. Rice\altaffilmark{1}, Jacqueline K. Faherty\altaffilmark{1,2,3}, Kelle L. Cruz\altaffilmark{1,4,5,6}}}

\altaffiltext{1}{Department of Astrophysics, American Museum of Natural History, New York, NY 10024, USA, erice@amnh.org}
\altaffiltext{2}{Department of Physics and Astronomy, Stony Brook University Stony Brook, NY 11794, USA}
\altaffiltext{3}{Visiting astronomer, Cerro Tololo Inter-American Observatory, National Optical Astronomy Observatory, which are operated by the Association of Universities for Research in Astronomy, under contract with the National Science Foundation.}
\altaffiltext{4}{Astronomy Department, California Institute of Technology, Pasadena, CA 91125, USA}
\altaffiltext{5}{Department of Physics and Astronomy, Hunter College, New York, NY 10065, USA}
\altaffiltext{6}{Visiting Astronomer at the Infrared Telescope Facility, which is operated by the University of Hawaii under Cooperative Agreement No. NCC 5-538 with the National Aeronautics and Space Administration, Science Mission Directorate, Planetary Astronomy Program}

\begin{abstract}
We present spectral and kinematic evidence that 2MASS~J06085283$-$2753583 (M8.5$\gamma$) is a member of the $\beta$ Pictoris Moving Group (BPMG, age~$\sim$12~Myr), making it the latest-type known member of this young, nearby association. We confirm low-gravity spectral morphology at both medium and high resolutions in the near-infrared. We present new radial velocity and proper motion measurements and use these to calculate galactic location and space motion consistent with other high-probability members of the BPMG. The predicted mass range consistent with the object's effective temperature, surface gravity, spectral type, and age is 15--35~$M_{Jup}$, placing 2MASS~0608$-$27 well within the brown dwarf mass regime. 2MASS~J06085283$-$2753583 is thus confidently added to the short list of very low mass, intermediate age benchmark objects that inform ongoing searches for the lowest-mass members of nearby young associations. 
\end{abstract}

\keywords{brown dwarfs --- infrared: stars --- open clusters and associations: individual ($\beta$ Pictoris) --- stars: individual (2MASS J06085283$-$2753583) --- stars: late-type --- techniques: spectroscopic}

\section{\sc Introduction}
Low mass members of nearby clusters, moving groups, and loose associations are important objects of study for stellar and planetary astrophysics, formation, and evolution. These groups are older than regions of ongoing star formation (e.g., in Taurus and Orion, ages $\sim$1--5~Myr) but younger than the disk population (ages $\sim$1--10~Gyr). Such associations have only recently been discovered because they are low-density groups and their members are spread out on the sky. Currently identified groups include the TW~Hydrae Association (TWA, age $\sim$10~Myr), the $\beta$ Pictoris Moving Group (BPMG, $\sim$12~Myr), and the Tucana/Horologium Association (Tuc-Hor, $\sim$30~Myr), along with several other clusters/associations with ages younger than $\sim$500~Myr \citep[e.g.,][]{Zuckerman04,Torres08}. 

Young low mass objects are hotter and have larger radii than older objects of the same mass \citep[e.g.,][]{Chabrier00}, thus they are brighter and can be studied more efficiently than their older counterparts. Young associations and moving groups are closer to the Sun than the closest star-forming regions ($d$$\sim$120--140~pc), allowing for more detailed study of their lowest-mass members. Furthermore, low mass members of slightly older associations ($>$20~Myr) have typically ceased accretion and dissipated their primordial disks, making it easier to disentangle atmospheric and circumstellar properties. On the other hand, they still may show signs of youth, such as low gravity spectral features and activity indicators. Eventually a complete census of the low mass members of nearby associations will address outstanding questions in star formation, including the low mass end of the initial mass function and its dependence on environment as well as the mass and age dependence of various processes in star and planet formation. 

This Letter establishes 2MASS~J06085283$-$2753583 (hereafter 2MASS~0608$-$27) as the lowest-mass confirmed member of the BPMG. Section~2 describes previous observations of 2MASS~0608$-$27, and Section~3 presents the observations analyzed herein. The kinematic and spatial properties of 2MASS~0608$-$27 are determined in Section~4, and its membership in the BPMG is discussed in Section~5. We present the conclusions of this Letter in Section~6.

\section{Previous Observations of 2MASS~0608$-$27}

2MASS~0608$-$27 was discovered by \citet{Cruz03} in a photometrically selected sample of ultracool dwarfs from the Two Micron All-Sky Survey (2MASS, \citealt{Skrutskie06}). They obtained a red-optical spectrum at 5.5~\AA~resolution with the RC~Spectrograph on the Blanco 4-m telescope on Cerro Tololo. In addition to H$\alpha$ emission they observed enhanced VO absorption, suggesting low gravity (their Figure~10). They did not assign a spectral type but compared 2MASS~0608$-$27 to standard M8 and M9 objects and estimated a rough distance of 30~pc.

\citet{Kirkpatrick08b} included 2MASS~0608$-$27 in their study of field late-M through L dwarfs exhibiting signatures of youth. They obtained red-optical spectra at 10~\AA~resolution with LRIS on Keck~I and noted weaker K~{\sc i}, Na~{\sc i}, and CaH and stronger VO than typically observed in field objects, indicative of low gravity \citep{McGovern04}, as well as H$\alpha$ emission (their Figure~8). They assign a spectral type of M8.5 `peculiar' and estimate the age of 2MASS~0608$-$27 to be $<$100~Myr based on comparison of the spectrum to the Pleiades late-M dwarf Teide~1.

\citet{Luhman09} reported  {\it Spitzer}/IRAC observations of 2MASS~0608$-$27 and found that its [3.6]$-$[5.8] and [3.6]$-$[8.0] colors showed no signs of infrared excess from a circumstellar disk. This generally suggests a lower age limit of 1--5~Myr, depending on the disk dissipation timescale for very low mass objects, although objects as young as 0.5~Myr have been observed without evidence for a disk.

\citet{Rice10} included 2MASS~0608$-$27 in their 5--10~Myr age group (also containing Upper Scorpius and TWA members) based on medium resolution ($R\sim$2000) $J$-band spectral features, namely weak K~{\sc i} lines \citep{McGovern05}. Spectral fitting with models confirmed the object's low gravity, consistent with an age of $\sim$5--10~Myr. We present a more detailed analysis of the observed $J$-band spectra in Section~\ref{nirspec}.

The proper motion of 2MASS~0608$-$27 was measured by \citet{Caballero07b} and \citet{Faherty09}. Both authors detected minor proper motion in right ascension ($\mu_{\alpha}$) and declination ($\mu_{\delta}$) with relatively large error bars: ($\mu_{\alpha}$, $\mu_{\delta}$) of (30$\pm$30, $-$30$\pm$30)~mas and ($-$13$\pm$11, $-$2$\pm$13)~mas, respectively. We present a new proper motion measurement in Section~\ref{PM}.

\section{\sc Near-Infrared Observations}
We analyzed near-infrared astrometric and spectroscopic observations of 2MASS~0608$-$27 to evaluate evidence of youth and membership in nearby stellar associations. Observations of 2MASS~0608$-$27 are summarized in Table~\ref{obslog} and described below.

\subsection{IRTF/SpeX $JHK$}
\label{spex}
2MASS~0608$-$27 was observed with SpeX on the NASA Infrared Telescope Facility \citep{Rayner03} as part of a larger investigation of field L dwarfs at low and moderate resolution in the near-infrared (K.~L.~Cruz et al. 2010, in preparation). The spectra were obtained in short-wavelength cross-dispersed mode, providing a wavelength coverage of 0.9--2.4~$\mu$m in six orders. Use of the 0.5\arcsec~slit provided a resolving power of $R\sim$1200. The A0~V star HD~52487 was observed  immediately following near a similar air mass as the target to provide both telluric correction and flux calibration. Data were reduced and corrected for telluric absorption using the SpeXtool package version 3.4 and standard methods \citep{Vacca03,Cushing04,Cushing05}.

\begin{figure}
  \includegraphics[height=.85\textheight,angle=90]{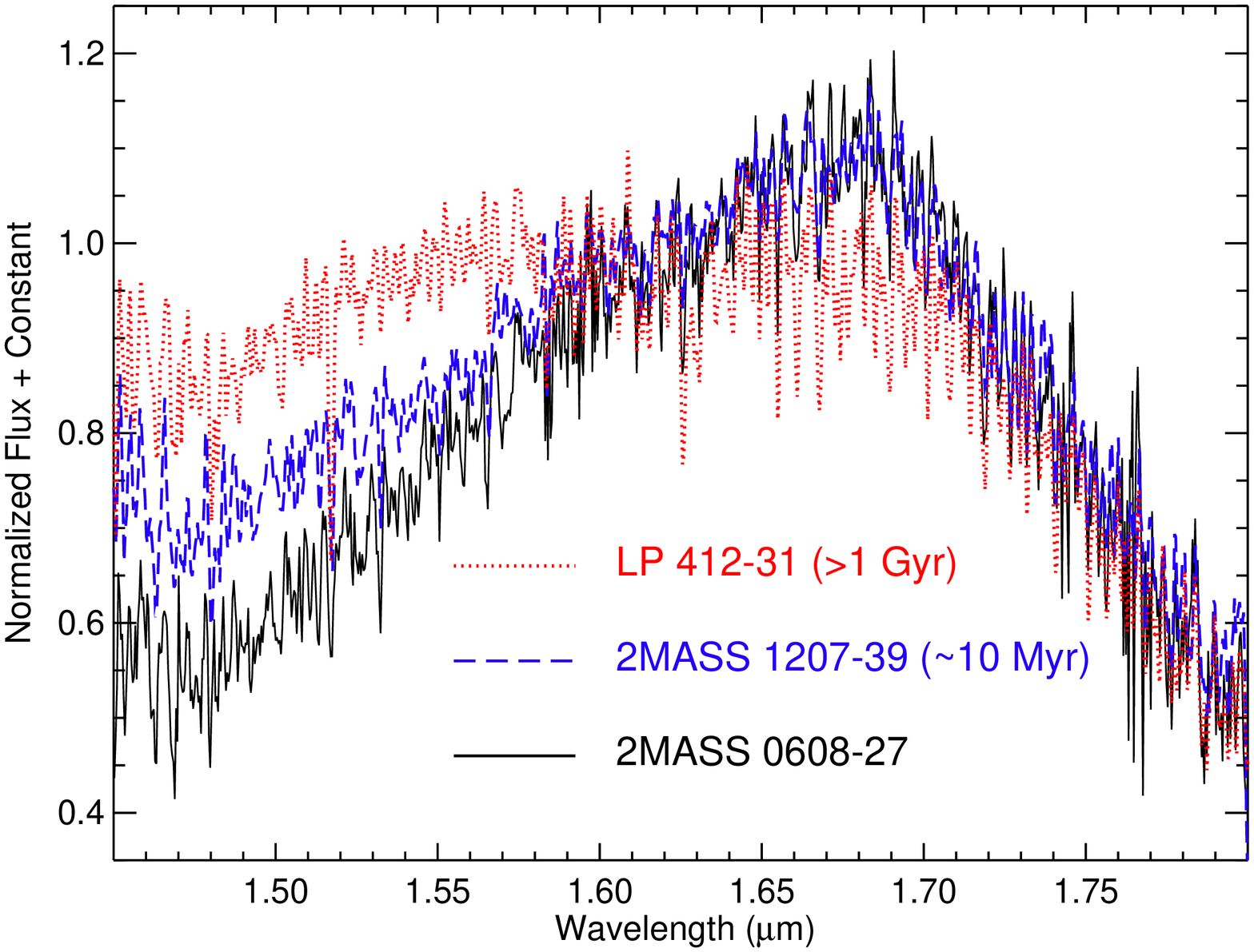}
  \caption{\label{h}SpeX cross-dispersed $H$-band spectra ($R\sim$1200) of 2MASS~0608$-$27 (solid black line) compared to spectra of the field M8 object LP~412-31 (dotted red line) and the young M8 TWA member 2MASS~1207$-$39 (dashed blue line) normalized at 1.8~$\mu$m. The highly peaked pseudo-continuum shape seen in 2MASS~1207$-$29 and 2MASS~0608$-$27 is typically interpreted as a hallmark of low gravity.
}
\end{figure}

In Figure~\ref{h} we compare the $H$-band SpeX spectrum with two late-type M dwarfs: the young M8 TWA member 2MASS~J12073347$-$3932540 (hereafter 2MASS 1207$-$39, data from K. Allers 2010, private communication) and the field M8 dwarf LP~412-31 \citep{Rayner09}. The spectrum of 2MASS~0608$-$27 shows a highly peaked $H$-band shape compared to the field dwarf LP~412-31, which is interpreted as stronger H$_2$O absorption on either side of the peak or weaker collisionally induced H$_2$ absorption. Similar spectral morphology has been observed in $\sim$1~Myr old objects \citep{Lucas01,Allers07} and M~giants \citep[e.g.,][Figure~105]{Rayner09} and is attributed to low surface gravity. 

\subsection{Keck~II/NIRSPEC $J$ Band}
\label{nirspec}
$J$-band spectra of 2MASS~0608$-$27 were obtained at both medium ($R$$\sim$2000) and high ($R$$\sim$20,000) resolutions with NIRSPEC on Keck~II \citep{McLean98,McLean00} as part of the Brown Dwarf Spectroscopic Survey (BDSS, \citealt{McLean03,McLean07}). The observing procedure and data reduction methods are summarized in \citet{Rice10}.

\begin{figure}
  \includegraphics[height=.85\textheight,angle=90]{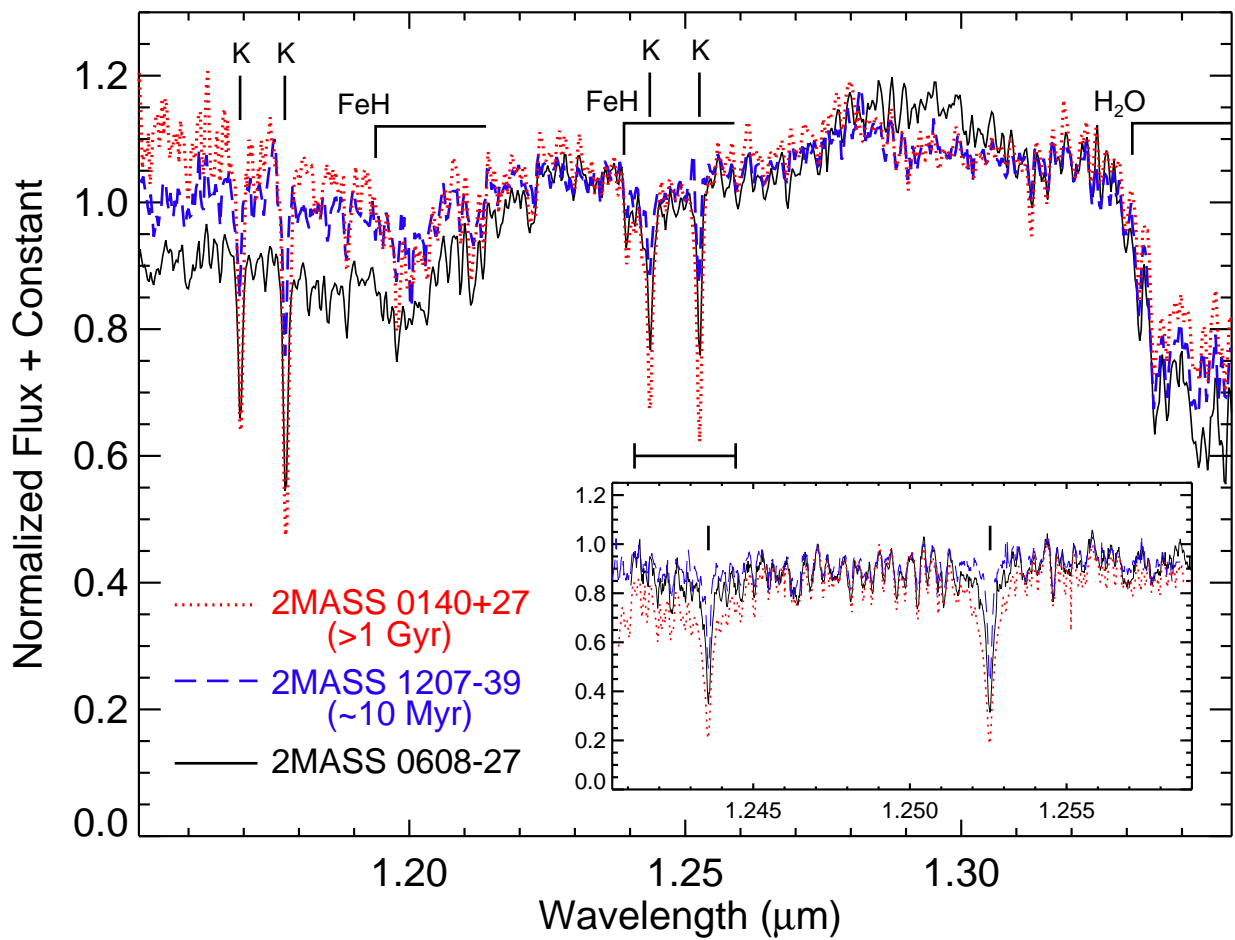}
  \caption{\label{J}NIRSPEC $J$-band spectra ($R\sim$2,000) of 2MASS 0608$-$17 (solid black line) compared to the young M8 TWA member 2MASS~1207$-$39 (dashed blue line) and the field M9 dwarf 2MASS~0140+27 (dotted red line) normalized at 1.25~$\mu$m. The inset plot shows the same objects observed at high resolution ($R\sim$20,000). The wavelength range of the inset plot is marked above it. 2MASS~0608$-$27 exhibits K~{\sc i} line strengths intermediate between those of the field and the young object, most apparent in the 1.25~$\mu$m doublet.
}
\end{figure}

Figure~\ref{J} shows spectra of 2MASS~0608$-$27 compared to a young M8 (2MASS~1207$-$39) and a field M9 (2MASS~J01400263+2701505, hereafter 2MASS~0140+27). The widths and depths of the K~{\sc i} lines of 2MASS~0608$-$27 are intermediate between those of the young M8 and the field M9. The K~{\sc i} lines have been shown to increase in strength with increasing gravity \citep{McGovern04,Allers07} and become pressure-broadened with both decreasing temperature and increasing surface gravity \citep{McLean07,Rice10}. Therefore the K~{\sc i} lines of 2MASS~0608$-$27 provide evidence for a cool temperature and low surface gravity. Interestingly the pseudo-continuum shape of 2MASS~0608$-$27 is redder than the spectra of both 2MASS~1207$-$39 and 2MASS~0140+27, it has excess flux at 1.28--1.30~$\mu$m, and the water band beginning at 1.33~$\mu$m is slightly stronger. This could indicate that 2MASS~0608$-$27 has a cooler temperature than implied by its red-optical spectral type or a dustier atmosphere than the comparison objects.

Additional $J$-band echelle orders of 2MASS~0608$-$27 from \citet{Rice10} and comparison spectra from the BDSS are used to measure radial velocity (RV, Section~4.2).

\subsection{Gemini/Phoenix $H$ Band}
Phoenix on the Gemini-South telescope was used to obtain a high-resolution ($R\sim$40,000) spectrum of 2MASS~0608$-$27 in the $H$ band (H6420 filter centered near 1.55~$\mu$m). The observation was made as part of a larger project to measure RVs of late-M and L spectral type objects in the field that show signatures of low gravity (K.~L.~Cruz et al. 2010, in preparation). The spectrum was extracted, wavelength calibrated (using OH emission lines), and flat-fielded using the IDL-based REDSPEC software\footnote{$http://www2.keck.hawaii.edu/inst/nirspec/redspec.html$}. Observations of two late-type objects with stable RVs were obtained with the same instrumental set-up for the cross-correlation. The RV measurements are discussed in Section~\ref{RV}.

\subsection{ISPI Imaging}
We observed 2MASS~0608$-$27 with the Infrared Side Port Imager (ISPI) on the 4-m Blanco telescope as part of an ongoing brown dwarf astrometric program (J.~K.~Faherty et al. 2010, in preparation). 2MASS~0608$-$27 was observed in the $J$-band filter for a total of 40~s at each of five dither positions. Individual point sources were selected in each reduced image and matched to stars in the 2MASS Point Source Catalog. Once an initial World Coordinate System (WCS) was set, the distortion across the ISPI detector was corrected. The final WCS residuals against 2MASS were $\sim$0.1~pixel (30~mas) in both $X$ and $Y$. Astrometry was performed on the final shifted and added image, and the results are described in Section~\ref{PM}. 

\section{\sc Kinematics and Distance}
Given its low gravity and implied youth, we calculate $UVW$ velocities and $XYZ$ location in order to test for membership in a young association.

\subsection{Proper Motion}
\label{PM}
We determined a new proper motion for 2MASS~0608$-$27 by combining the RA and DEC position from ISPI with the 2MASS position. We calculated ($\mu_{\alpha}$, $\mu_{\delta}$) to be ($-$5$\pm$ 5, 10$\pm$5)~mas~yr$^{-1}$ using the 10.9 yr baseline between images (Table~\ref{objprop}). Our measurement is consistent with previous measurements of \citet{Caballero07b} and \citet{Faherty09} with refined uncertainties from the longer time baseline of our observations. 

\subsection{Radial Velocity}
\label{RV}
We measure the RV of 2MASS~0608$-$27 by cross-correlating the NIRSPEC echelle order 62 spectrum with multi-epoch spectra of RV standards Gl~406 (M6, \citealt{Nidever02}) and 2MASS~0140+27 (M9, \citealt{Reid02}) from the BDSS. The mean measurement is 23.4$\pm$0.5~km~s$^{-1}$. Cross-correlation of the reference spectra against one another provide an estimate of 1.0~km~s$^{-1}$ for the systematic uncertainty. We also measured the RV of 2MASS~0608$-$27 using the high-resolution ($R$$\sim$40,000) Phoenix $H$-band spectrum via cross correlation with spectra of two L dwarfs from \citet{Blake07}, 2MASS~J05233822$-$1403022 and 2MASS~J11553952$-$3727350. The result was 24.5$\pm$1.5~km~s$^{-1}$. We adopt the average of these measurements, RV=24.0$\pm$1.0~km~s$^{-1}$. This value is consistent with the previous measurement of RV=22.6$\pm$1.5~km~s$^{-1}$ by \citet{Rice10}.

\subsection{Distance Estimate}
The distance to 2MASS~0608$-$27 was estimated by \citet{Faherty09} to be 27$\pm$4~pc, based on the absolute 2MASS $J$-band magnitude and spectral type relation in \citet{Cruz03}.  However young objects at a given spectral type are typically brighter than their old counterparts. Indeed, initial parallax results for low surface gravity dwarfs demonstrate that distances are underestimated by up to 10$\%$ (J.~K.~Faherty et al. 2010, in preparation).  Therefore we adopt a distance of 30~pc and increase the estimated uncertainty to $\pm$10~pc.

\section{\sc Discussion}
\subsection{The $\beta$ Pictoris Moving Group}
Shortly after \citet{BarNav99} searched for kinematic companions to the young star $\beta$~Pictoris in order to further constrain its age, \citet{Zuckerman01} proposed 27 co-moving stars as the BPMG. Proposed members were required to show some signature of youth, for example: large fractional X-ray luminosity, rapid rotation, large lithium abundance, or H$\alpha$ emission. The age of the BPMG is typically quoted as $\sim$12~Myr, which is consistent with age estimates from dynamical back-tracking models, age estimates for individual objects, and the distribution of lithium abundance \citep[][and references therein]{Fernandez08,Torres08}. New members have been identified by several authors: \citet{Torres08} included 48 stars as high-probability BPMG members, and \citet{Lepine09} added three members and one common proper motion companion. The low-mass membership of the BPMG is certainly incomplete. 

 The high-probability members of the BPMG listed by \citet{Torres08} include 21 objects with spectral types from M0 to M5, and the latest type objects, three M4 and two M5 objects, are all in multiple systems.Other low mass objects in the BPMG include the companion to $\eta$~Tel (HR~7329), an M7.5 brown dwarf \citep{Lowrance00}. In addition, \citet{Reiners09b} and \citet{Reiners10} determined that the young M6.5/M9 binary 2MASS J0041353$-$562112 could be an ejected member of BPMG, although they state that it is more likely a member of the $\sim$20~Myr old Tuc-Hor Association. $\beta$~Pictoris itself has a candidate planetary-mass companion, although the object has not been detected since the discovery observations \citep{Lagrange09a,Lagrange09b,Fitz09}.

\subsection{Space Motion and Location}
\begin{figure}
  \includegraphics[height=.85\textheight,angle=90]{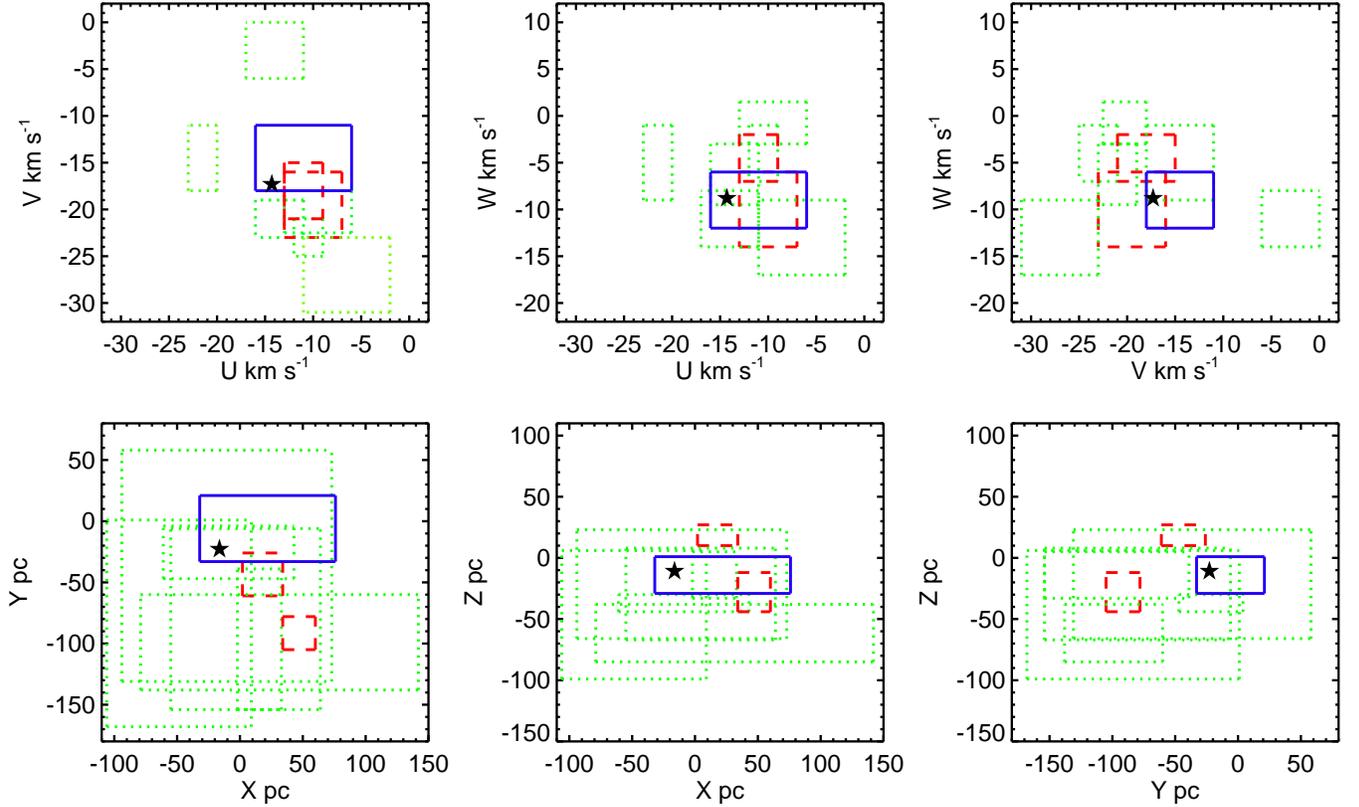}
  \caption{\label{uvwxyz}$UVWXYZ$ properties of 2MASS~0608$-$27 (filled star) compared to those of nearby young associations presented in several sub-spaces. Top row: $UVW$ velocity sub-space plots, bottom row: $XYZ$ location sub-space plots. Rectangles surround the furthest extent of highly probable members from \citet{Torres08}, but their distribution does not necessarily fill the entire rectangle. Dashed lines (red) represent the younger moving groups TWA ($\sim$10~Myr) and $\epsilon$~Cha ($\sim$6~Myr), the solid line (blue) represents the BPMG, and dotted lines (green) represent the older associations (Tuc-Hor, Columba, Carina, Octans, Argus, and AB~Doradus) with ages of $\sim$20--70~Myr. The space velocity and location of 2MASS~0608$-$27 are only coincident with the BPMG in all sub-spaces.
}
\end{figure}

Based on the measured proper motion and RV and the distance estimate described above, we calculate the $UVW$ space motion of 2MASS~0608$-$27 (in a left-handed coordinate system with $U$ positive toward the Galactic center) to be ($-$14.3$\pm$1.4, $-$17.3$\pm$1.3, $-$8.8$\pm$1.4)~km~s$^{-1}$. The $XYZ$ coordinates of 2MASS~0608$-$27 calculated from its RA, DEC, and distance are ($-$16.3$\pm$5.4, $-$22.8$\pm$7.6, $-$10.8$\pm$3.6)~pc. The $UVWXYZ$ of 2MASS~0608$-$27 are shown relative to the loose associations from \citet{Torres08} in Figure~\ref{uvwxyz}. Uncertainties on $UVWXYZ$ values were estimated using Monte Carlo methods and are listed in Table~\ref{objprop}. The results are most consistent with the BPMG, which has a mean $UVW$ of ($-$10.1$\pm$2.1, $-$15.9$\pm$0.8, $-$9.2$\pm$1.0)~km~s$^{-1}$ \citep{Torres08}. 2MASS~0608$-$27 is within 2$\sigma$ of the mean space velocity of high probability BPMG members. The $XYZ$ coordinates of confirmed BPMG members are centered around (20$\pm$50, $-$5$\pm$25, $-$15$\pm$14)~pc \citep[][Table~2 and Figure~8]{Torres08}.

\subsection{Membership of 2MASS~0608$-$27 in the BPMG}
The spectral similarities of 2MASS~0608$-$27 and the TWA member 2MASS~1207$-$39 support a young age for 2MASS~0608$-$27, but low-mass objects vary drastically in their exhibition of youth signatures like X-ray emission and accretion indicators. The membership of 2MASS~0608$-$27 in the BPMG provides a tighter age constraint because a reliable age for the cluster has been determined with consistent results from different methods. Figure~\ref{uvwxyz} shows the $UVWXYZ$ distributions of young loose associations from \citet{Torres08} in six sub-spaces; 2MASS~0608$-$27 is only coincident with the BPMG in all sub-spaces.

\subsection{Physical Properties of 2MASS~0608$-$27}
\label{props}
We revisit the spectral type estimation of \citet{Cruz03} and \citet{Kirkpatrick08b} using the methods of \citet{Cruz09} and assign a type of M8.5$\gamma$ to 2MASS~0608$-$27, with the $\gamma$ designation indicating low gravity spectral features in the optical, intermediate between those of Pleiades members (designated $\beta$) and Taurus members ($\delta$). This is supported by the overall spectral similarity to the young M8 TWA member 2MASS~1207$-$39 in the near-infrared (Figures~\ref{h} and \ref{J}), but deeper H$_2$O absorption and a redder $J$-band continuum could indicate a lower temperature and/or a dustier atmosphere for 2MASS~0608$-$27 (Section~3.1 and 3.2).

\citet[][their Figure~11]{Rice10} compared the effective temperature and surface gravity of 2MASS~0608$-$27 inferred from spectral fitting to DUSTY00 evolutionary models \citep{Chabrier00}. The result supports an age of $\sim$9~Myr and a mass of $\sim$25~$M_{Jup}$ for 2MASS~0608$-$27. It is noteworthy that the \citet{Rice10} high- and medium-resolution fits produced considerably different effective temperatures, 2529~K and 2200~K, respectively. This indicates the continued difficulty of inferring effective temperatures of cool objects from model fits. However, at the surface gravity inferred from the high-resolution fit, the implied age is relatively insensitive to temperature. Even an effective temperature of 2200~K would still imply an age older than 5~Myr with a mass above $\sim$15~$M_{Jup}$, which we therefore include in the range of possible masses. The physical properties of 2MASS~0608$-$27 can also be inferred from its similarity to young TWA members 2MASS~1207$-$39 which has an estimated mass of 25$\pm$5~$M_{Jup}$ using photometry, trigonometric parallaxes, an age implied by TWA membership, and the DUSTY00 evolutionary models \citep{Teixeira08}. As a member of the BPMG, 2MASS~0608$-$27 would be slightly older than the TWA members, which we account for with a larger range of masses than are typically quoted for 2MASS~1207$-$39. We therefore estimate a mass of 15--35~$M_{Jup}$. The physical properties of 2MASS~0608$-$27 are assembled in Table~\ref{objprop}.

\section{\sc Conclusions}
Spectral features, kinematics, and the space location of 2MASS~0608$-$27 clearly confirm it as a member of the BPMG. 2MASS~0608$-$27 is the latest-type non-companion member by a substantial margin. Similarly late-type objects in the BPMG are the M7.5 companion HR~7329B, possibly the M6.5/M9 binary 2MASS~J0041353$-$562112, and the candidate planetary-mass companion to $\beta$~Pictoris. Ongoing proper motion and RV measurements for apparently young M and L dwarfs in the field will likely confirm larger numbers of very low mass members of the BPMG and other young, nearby loose associations (\citealt{Schlieder10}, K.~L.~Cruz et al. 2010, in preparation). The identification of the M8.5 object 2MASS~0608$-$27 as a member of the BPMG provides an important late-type benchmark object for further studies.

\section{Acknowledgments}
The authors thank K.~Allers for her comments on an early draft of this letter and J.~D.~Kirkpatrick, M.~Cushing, M.~Simon, J.~Schlieder, and S.~Metchev for useful discussions and the anonymous referee for helpful comments. Observing and/or data reduction assistance were provided by M.~R. McGovern and G.~Gimeno. Support for this work was provided by NASA through the $Spitzer Space Telescope$ Fellowship Program, through a contract issued by JPL/Caltech. The authors wish to acknowledge the staff of the Keck, Cerro Tololo, and IRTF observatories for their outstanding support.

Data presented herein were obtained at the W.~M. Keck Observatory, which is operated as a scientific partnership among the California Institute of Technology, the University of California and the National Aeronautics and Space Administration. The observatory was made possible by the generous financial support of the W.~M.~Keck Foundation. This research has made use of the NASA/IPAC Infrared Science Archive operated by the Jet Propulsion Laboratory, California Institute of Technology, under contract with the National Aeronautics and Space Administration. This publication makes use of data from the Two Micron All Sky Survey, which is a joint project of the University of Massachusetts and the Infrared Processing and Analysis Center, funded by the National Aeronautics and Space Administration and the National Science Foundation. This research has made use of the SIMBAD database, operated at CDS, Strasbourg, France and NASA's Astrophysics Data System. Finally, the authors wish to extend special thanks to those of Hawaiian ancestry on whose sacred mountain we are privileged to be guests.

{\it Facilities:} \facility{Keck:II (NIRSPEC)}, \facility{IRTF (SpeX)}, \facility{Blanco (ISPI)}, \facility{Gemini-South (Phoenix)}

\bibliography{/Users/erice/Documents/AMNH/2m0608/refs}

\clearpage

\begin{deluxetable}{llcrrc}
\tablewidth{0pt} 
\tablecaption{\bf Observing log of 2MASS J06085283$-$2753583
\label{obslog} }
\tablehead{ 
\colhead{UT Date} &
\colhead{Telescope/} &
\colhead{Band} &
\colhead{Resolving} &
\colhead{Integration} &
\colhead{Reference }\\
\colhead{of Observation} &
\colhead{Instrument } &
\colhead{} &
\colhead{Power} &
\colhead{Time (s)} &
\colhead{} }
\startdata
2003 Dec 4  & Keck~II/NIRSPEC  & $J$   & 2000~~~   &  600~~~~ & 1, 2 \\
2007 Nov 13 & IRTF/SpeX        & $JHK$ & 1200~~~   & 1600~~~~ & 3 \\
2008 Dec 6  & Keck~II/NIRSPEC  & $J$   & 20000~~~ & 4800~~~~ & 2 \\
2009 Oct 28 & Gemini-S/Phoenix & $H$   & 40000~~~ & 2400~~~~ & 3 \\
2009 Dec 12 & CTIO/ISPI        & $J$   &Imaging~~~ & 200~~~~  & 4 \\
\enddata
\tablerefs{(1) \citet{McGovern05}; (2) \citet{Rice10}; (3) this work; (4) J.~K.~Faherty et al. 2010, in preparation. } 
\end{deluxetable}

\begin{deluxetable}{lrc}
\tablewidth{0pt} 
\tablecaption{\bf Properties of 2MASS J06085283$-$2753583 
\label{objprop}}
\tablehead{ 
\colhead{Parameter} &
\colhead{Value} &
\colhead{Reference} } 
\startdata
RA, DEC (J2000)     & 06$^h$ 08$^m$ 52.8$^s$, $-$27\degr~53\arcmin~58\arcsec         & 1 \\
Optical spectral type  & M8.5$\gamma$                   & 2 \\
$J$, $H$, $K$    & 13.595$\pm$0.028, 12.897$\pm$0.026, 12.371$\pm$0.026  & 1 \\
$\mu$$_{\alpha}$, $\mu$$_{\delta}$ & $-$5$\pm$5 mas yr$^{-1}$, 10$\pm$5 mas yr$^{-1}$ & 2 \\
RV               & 24.0$\pm$1.0 km~s$^{-1}$     & 2 \\
$d$$_{est}$      & 30$\pm$10 pc               & 2 \\
$UVW$              &$-$14.3$\pm$1.4, $-$17.3$\pm$1.3, $-$8.8$\pm$1.4~km~s$^{-1}$    & 2 \\
$XYZ$              &$-$16.3$\pm$5.4, $-$22.8$\pm$7.6, $-$10.8$\pm$3.6~pc      & 2 \\
$T_{eff}$        & 2555~K, 2529~K\tablenotemark{a}     & 3, 4 \\
log(g)           & 3.98                   & 4 \\
Age              & $\sim$10~Myr               & 2 \\
Mass             & 15--35~$M_{Jup}$  & 2 \\
\enddata
\tablerefs{(1)~2MASS~\citep{Skrutskie06}; (2)~this work; (3)~\citet{Luhman03}; (4)~\citet{Rice10}. }
\tablenotetext{a}{The medium-resolution $J$-band spectrum of 2MASS~0608$-$27 is best fit by $T_{eff}$=2200~K \citep{Rice10}, which is accounted for in the range of masses estimated here (see Section~\ref{props}).}
\end{deluxetable}

\end{document}